\documentclass[twocolumn,showpacs,amsmath,amstex,amssymb,mathfonts,prb]{revtex4-1}
\usepackage{graphicx,bm,units,verbatim,color,times}

\begin{document}

\title{Fractional quantum Hall effect in a tilted magnetic field}
\author{Z. Papi\'c}
\affiliation{Department of Electrical Engineering, Princeton University, Princeton, NJ 08544, USA}

\date{\today}

\begin{abstract} 
We discuss the orbital effect of a tilted magnetic field on the quantum Hall effect in parabolic quantum wells. Many-body states realized at the fractional 1/3 and 1/2 filling of the second electronic subband are studied using finite-size exact diagonalization. In both cases, we obtain the phase diagram consisting of a fractional quantum Hall fluid phase that persists for moderate tilts, and eventually undergoes a direct transition to the stripe phase. It is shown that tilting of the field probes the geometrical degree of freedom of fractional quantum Hall fluids, and can be partly related to the effect of band-mass anisotropy. 
\end{abstract}


\maketitle \vskip2pc

\section{Introduction}\label{sec_introduction}

When a thin 2D layer of highly mobile charge carriers is placed in a perpendicular magnetic field, it gives rise to a fascinating variety of phases~\cite{tsg} that have been the subject of intensive study over the last three decades~\cite{prange}. Among the most remarkable of these phases are those that display an excitation gap, and therefore quantized Hall conductivity in transport, even when their valence Landau level is only partially filled~\cite{tsg}. This phenomenon -- the fractional quantum Hall effect (FQHE) -- was soon realized to be a direct manifestation of the many-body nature of these phases, and a variety of elaborate theoretical concepts have been put forward to understand it. These techniques have included, \emph{inter alia}, the method of writing down ``inspired" first-quantized wavefunctions \`a la Laughlin~\cite{laughlin} which have been shown to possess intricate analytical structure~\cite{jack}; the Chern-Simons topological field theory~\cite{zhk} and the conformal field theory~\cite{mr}; composite fermion theory~\cite{jainbook}; and of course the explicit microscopic calculations based on exact-diagonalization~\cite{ed}. Perhaps somewhat surprisingly, the last method has been particularly successful due to the specific nature of the correlations in FQHE that rapidly quench the finite-size effects as the number of particles is increased. A most striking example of this occurs for the fractionally filled $\nu=1/3$ Laughlin state, where the essential physical properties (the quantum numbers of the ground states, the type of collective excitation mode and its gap) can be identified in systems as small as 4 particles. Similarly, composite fermion theory~\cite{jainbook} in many cases achieves an astonishing quantitative accuracy in calculations of the overlaps between the composite-fermion trial wavefunctions and the exact ground states of the realistic Coulomb Hamiltonians.

Because of the synergy of the theoretical approaches mentioned above, a fairly good agreement between experiment and theory has been established in a number of cases. The level of agreement appears to be the best in the lowest Landau level (LL) defined by the filling factors $0<\nu<2$. In this range of $\nu$'s, the ``hierarchy" theory~\cite{haldane_sphere,haldane_prange,halperin} and composite-fermion theory~\cite{jainbook} account for nearly all of the observed experimental phenomenology. Because of the high magnetic field, in typical samples the role of multicomponent degrees of freedom, such as spin, is not crucial. However, recently FQHE also been observed in graphene~\cite{fqhe_graphene} where multicomponent degrees of freedom are known to play a much more subtle role~\cite{graphene_multicomponent}, and agreement between theory and experiment is generally poorer at this stage. 

Another effect that is potentially important is the so-called ``Landau-level mixing"~\cite{llmix,zed}. In most theoretical descriptions of a partially filled Landau level, excitations to other Landau levels are disregarded. This approximation is particularly desirable in numerics, where the inclusion of multiple Landau levels leads to an extremely rapid increase of the size of the Hilbert space. While physically the neglect of LL mixing appears to be reasonable in the lowest $n=0$ LL, recent work~\cite{llmix} shows that this approximation is somewhat poor for $n=1$ LL. This is unfortunate because some of the most exciting FQH states are realized in $n=1$ LL, such as the celebrated $\nu=5/2$~\cite{52} and $\nu=12/5$~\cite{125} states that are believed to possess non-Abelian quasiparticles in the bulk~\cite{mr,rr}. In the case of half filling, LL mixing is the mechanism that breaks the particle-hole symmetry, and selects one of the two candidate wavefunctions proposed for it~\cite{mr,antipf,llmix_shayegan,llmix_csathy}. More accurate treatment of LL mixing has inspired some recent work~\cite{llmix,zed}, but in general its effects on various filling factors are still poorly understood. 

\begin{figure*}[htb]
  \begin{minipage}[l]{\linewidth}
    \includegraphics[scale=2.2]{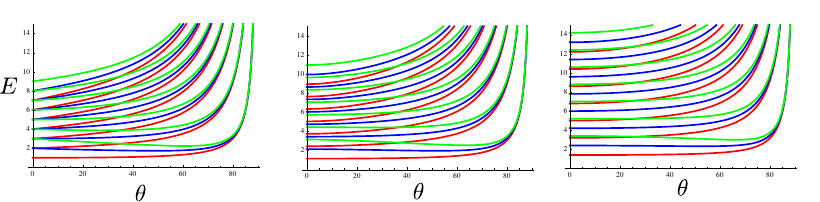}
  \end{minipage}
\caption{(color online) Energy levels $E=(n_1+1/2)\omega_1 + (n_2+1/2)\omega_2$ in units of $\omega_z$ as a function of tilt angle $\theta$, for several low-lying subbands $(n_1,n_2)$ and three choices of the confining potential $\omega_0/\omega_z=1, 1.3, 1.8$ (from left to right). Red lines correspond to the subbands $(0,n_2)$, blue lines correspond to $(1,n_2)$, green lines correspond to $(2,n_2)$. If the confinement is equal to 1, subbands $(0,1)$ and $(1,0)$ are degenerate, as expected.}
\label{fig_tilt}
\vspace{-0pt}
\end{figure*}

In this paper we discuss another mechanism that is also ubiquitous to many FQHE experiments, yet has not been fully understood theoretically. We consider the so-called ``tilted field" setup~\cite{tilt_old} where, in addition to the perpendicular component of the magnetic field along the $z$-axis, a parallel component of the field is introduced along $x$-axis. The total magnetic field in that case points along an angle $\theta$ with respect to the vertical axis, which we refer to as the tilt angle. This technique has been immensely popular in experiment~\cite{tilt_old, engel, dean, csathy_tilt, du, xia_tilt, xia2011, tsui_52, tsui_125, shayegan_tilt}, e.g. as a probe for spin polarization: to the lowest order, the only effect of parallel field is to increase the Zeeman coupling, therefore it distinguishes a polarized ground state from an unpolarized one. However, this is true only in the limit of a sample with zero thickness. Since the real samples are typically a few tens of nanometers thick, and the  experiments are often performed at very large tilt angles reaching 70-80 degrees, one expects a strong coupling between the tilt and the orbital motion of electrons. Therefore, the consequences of tilting could be far more dramatic for the many-body states than it naively appears. This was vividly illustrated in several recent experiments~\cite{dean, du, xia_tilt, xia2011, tsui_52, tsui_125, shayegan_tilt}.  

The remainder of this paper is organized as follows. In Sec.~\ref{sec_onebody} we review the solution of the one-body problem in a parabolic quantum well subject to a combination of perpendicular and parallel field. In Sec.~\ref{sec_manybody} we derive the matrix elements of the interaction Hamiltonian adapted for finite-size studies. In the main Sec.~\ref{sec_results} we present the results of numerical simulations for the fractional fillings 1/3 and 1/2 of one of the excited subbands that was the subject of recent experiments~\cite{xia2011,shayegan_tilt}. We conclude with the discussion of these results (Sec.~\ref{sec_conc}), pointing out their limitations, connection to related recent work in the literature, and future directions.

\section{One-body problem}\label{sec_onebody}

In this Section we review the quantum-mechanical solution for a single particle inside a parabolic quantum well confined in a potential $U=\frac{1}{2}m^* \omega_0^2 z^2$, with a perpendicular component of the magnetic field $B_z \hat z$ and a parallel ($x$) component $B_{||}$. This solution was first given in Maan \emph{et al.}~\cite{maan} (see also Refs.~\onlinecite{chak, demler}).
We consider parabolic confinement for simplicity, since an analytic solution is not possible for square quantum wells. However, the two models, square and parabolic, are expected to show the same qualitative features up to some rescaling of the effective width of the well (see Ref.~\onlinecite{peterson} for details).    

The input parameters are the tilting angle $\theta$, defined by $\tan \theta=B_{||}/B_z$, and $\omega_0/\omega_z$, where $\omega_0$ is the confinement frequency. We also have $\omega_c^2=\omega_x^2+\omega_z^2$ i.e. $\omega_z=\omega_c \cos \theta$, $\omega_x=\omega_c \sin \theta$. We are interested in preserving the filling factor as the field is tilted, i.e. we assume $\omega_z$ is constant (perpendicular field determines the filling factor), and hence we can work in units $\ell_0^2=\hbar/m\omega_z=1$. 

The one-body problem is conveniently solved in the Landau gauge $(0,B_z x-B_{||} z,0)$, and maps to a sum of two harmonic oscillators by the rotation in $x'-z$ plane, where $x'\equiv x+c k_y/e B_z$ and $k_y$ is the momentum along $y$ axis. The rotation angle $\phi$ is given by
$$
\tan 2\phi = \frac{-2\omega_x\omega_z}{\omega_0^2+\omega_x^2-\omega_z^2},
$$
and the frequencies of the two oscillators are given by
\begin{eqnarray}
\nonumber \omega_1^2=(\omega_0^2+\omega_x^2)\sin^2 \phi + \omega_z^2 \cos^2\phi + 2 \omega_x \omega_z \sin\phi \cos\phi, \\
\nonumber \omega_2^2=(\omega_0^2+\omega_x^2)\cos^2 \phi + \omega_z^2 \sin^2\phi - 2 \omega_x \omega_z \sin\phi \cos\phi. 
\end{eqnarray}
These frequencies define two effective magnetic lengths
$$
\ell_{1}^2=\omega_z/\omega_1,\ell_{2}^2=\omega_z/\omega_2.
$$
Note that $\omega_1$ denotes the frequency of the oscillator with the coordinate $x'\cos\phi - z \sin\phi$. Therefore, when the tilt angle is zero, $\omega_1 \to \omega_z$, and $\ell_1 \to 1$, which ensures that the problem correctly reduces to the situation without the parallel field. For $\phi=0$, the parabolic confinement has an effect only on the $z$ coordinate that couples to the magnetic length $\ell_2$ which may be different from $\ell_0$. 

Energy levels of a single particle are thus labelled by $(n_1,n_2)$ corresponding to the quantum numbers of the two oscillators. It is instructive to analyze the first few low-lying levels as a function of the tilt, as shown in Fig.~\ref{fig_tilt}. Energies are quoted in units of $\omega_z$ which is assumed to be constant.  In general, the subbands display a number of crossings as the parallel field is increased. However, we will focus on the two lowest subbands $(0,0)$ and $(1,0)$, which are experimentally relevant and do not cross any other subband or each other. In Fig.~\ref{fig_tilt} we show energy levels for three choices of confinement $\omega_0/\omega_z \geq 1$. These values correspond to higher mobility quantum wells, which are relevant for FQHE that we study below (in principle, $\omega_0/\omega_z <1$ is also possible but those quantum wells typically have lower mobilities). Note that for very large tilts (60 degrees and higher), the levels organize into subbands $(n_1,0)$,$(n_1,1)$, $(n_1,2)$ etc. In this interesting regime, each of the new emergent bands represents a continuum of LLs that have collapsed on top of each other. This strong-mixing regime is difficult to study theoretically, and the results of this paper are not expected to hold there. We will further comment on this in Sec.~\ref{sec_conc}.

In the Landau gauge with an open boundary condition along $x$ (cylinder geometry), the single-particle wavefunctions are given by 
\begin{eqnarray}
\nonumber \phi_{j}^{00}(x,y,z) &=& \frac{1}{\sqrt{b\pi \ell_1 \ell_2}} e^{i X_j y} \chi_{\ell_1} (\mathcal{X}) \chi_{\ell_2} (\mathcal{Z}), \\
\nonumber \phi_{j}^{10} (x,y,z) &=& \frac{1}{\sqrt{2b\pi \ell_1 \ell_2}} H_1 ( \frac{\mathcal{X}}{\ell_1} ) e^{i X_j y} 
\chi_{\ell_1} (\mathcal{X}) \chi_{\ell_2} (\mathcal{Z}), 
\end{eqnarray}
where $\chi_\ell(x)=\exp(-x^2/2\ell^2)$, $\mathcal{X}\equiv (X_j +x)\cos\phi - z \sin\phi$, $\mathcal{Z}\equiv (X_j +x)\sin\phi + z \cos\phi$, and $H_1$ is the first Hermite polynomial. The repeat distance along $y$ is denoted by $b$ and $X_j \equiv 2\pi j/b$, where $j$ is an integer labeling the orbitals. In order to eliminate edge effects, it is useful to consider a fully periodic boundary condition i.e. wrap the cylinder along $x$ onto a torus. The wavefunctions in that case are given by
\begin{eqnarray}\label{toruswf}
\Phi_{j}^{\sigma\sigma'} (x,y,z) &=& \sum_{k\in \mathbb{Z}} \phi_{j+kN_\phi}^{\sigma\sigma'} (x,y,z),
\end{eqnarray}
where $a$ is the dimension of torus along $x$, and the quantization of flux leads to the constraint $ab=2\pi N_{\phi}$ ($\sigma,\sigma'$ label the subbands). The index $j$ runs from 0 to $N_\phi-1$. The wavefunctions above are written for a rectangular torus; at the cost of a few extra complications, they can be straightforwardly generalized to an arbitrary twisted torus, which is needed if one wishes to study, e.g., a unit cell with highest (hexagonal) symmetry in the twodimensional plane.

\section{Many-body Hamiltonian}\label{sec_manybody}

Using the one-body wavefunctions in Eq.~(\ref{toruswf}), we can construct the interacting Hamiltonian by computing the matrix elements~\cite{yhl}:
\begin{eqnarray}\label{coulomb}
\nonumber \int d^2\mathbf{r}_1 d^2\mathbf{r}_2 \int dz_1 dz_2 \Phi_{j_1}^{\sigma_1\sigma_1'*}(\mathbf{r}_1,z_1) \Phi_{j_2}^{\sigma_2\sigma_2'*}(\mathbf{r}_2,z_2) \\
 V(\mathbf{r}_1-\mathbf{r}_2,z_1-z_2) \Phi_{j_3}^{\sigma_3\sigma_3'}(\mathbf{r}_2,z_2) \Phi_{j_4}^{\sigma_4\sigma_4'}(\mathbf{r}_1,z_1) 
\end{eqnarray}
where $\mathbf{r}$ denotes the vector in $x-y$ plane. As it stands, Coulomb interaction in Eq.~(\ref{coulomb}) admits all types of scattering processes from subbands $(\sigma_1\sigma_1'),(\sigma_2\sigma_2')$ into subbands $(\sigma_3\sigma_3'), (\sigma_4\sigma_4')$, subject to the momentum conservation (see Eq.(\ref{matel}) below). However, as we are mainly interested in partially filled (0,0) and (1,0) subbands, which become well-separated from other subbands for large values of the confinement (see Fig.~\ref{fig_tilt}), we will neglect all scattering processes between different subbands, i.e. retain only $\sigma_1=\ldots=\sigma_4$ and $\sigma_1'=\ldots=\sigma_4'$. This method is analogous to the ``lowest Landau level projection" commonly used in FQH finite size studies~\cite{prange}. We expect this approximation to become increasingly better as the confinement is increased.

\begin{figure}[htb]
\includegraphics[scale=0.35,angle=0]{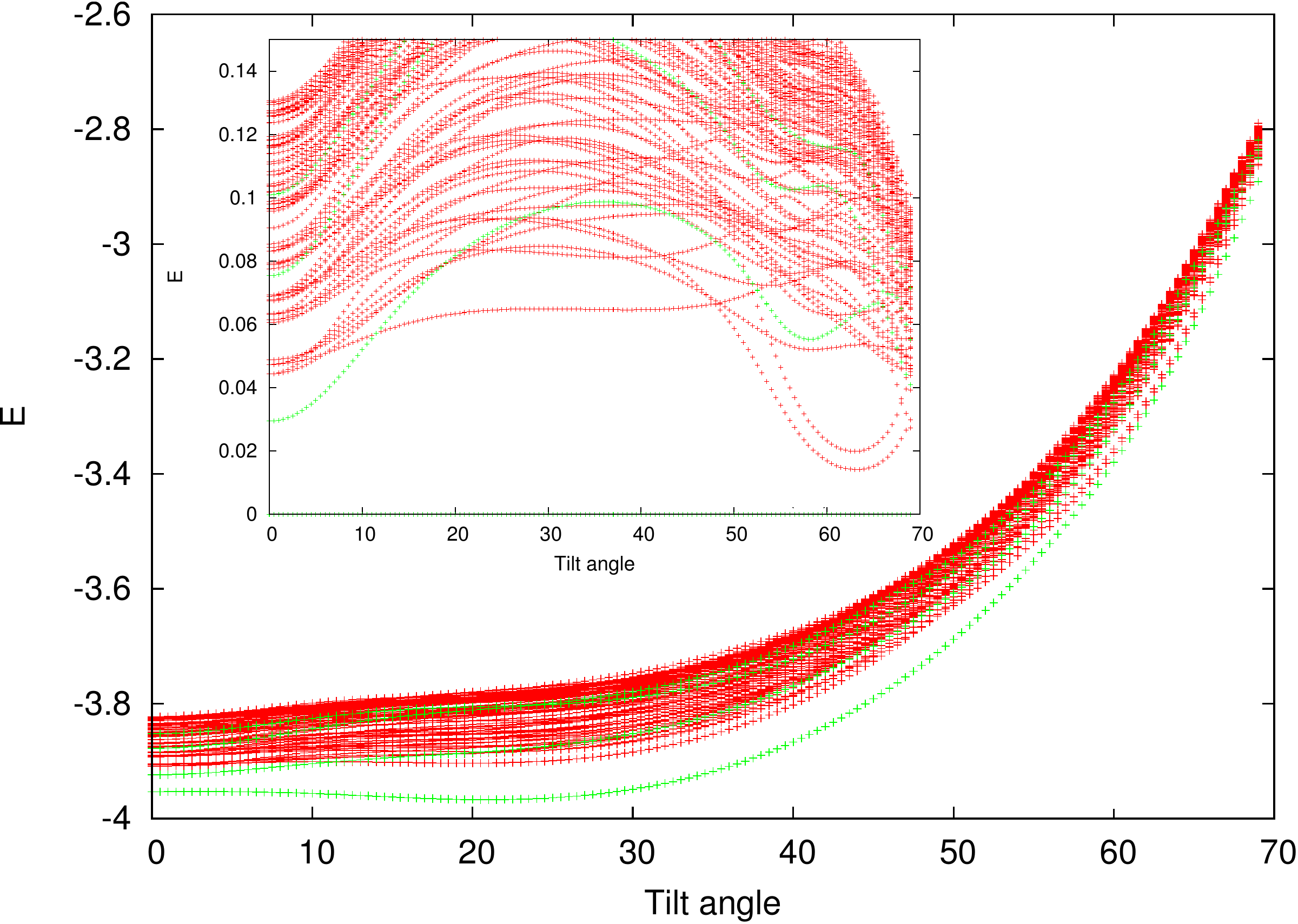}
\caption{(color online) Energy spectrum for $N=10$ particles at filling 1/3 of the subband (1,0) as a function of tilt angle (measured in degrees). Inset shows the same spectrum but plotted relative to the ground state at each tilt angle, which represents the neutral gap of the system as a function of tilt. Energy levels belonging to $\mathbf{k}=0$ momentum sectors are shown in green color.}
\label{fig_onethird}
\end{figure}

\begin{figure*}[ttt]
  \begin{minipage}[l]{\linewidth}
    \includegraphics[scale=0.8]{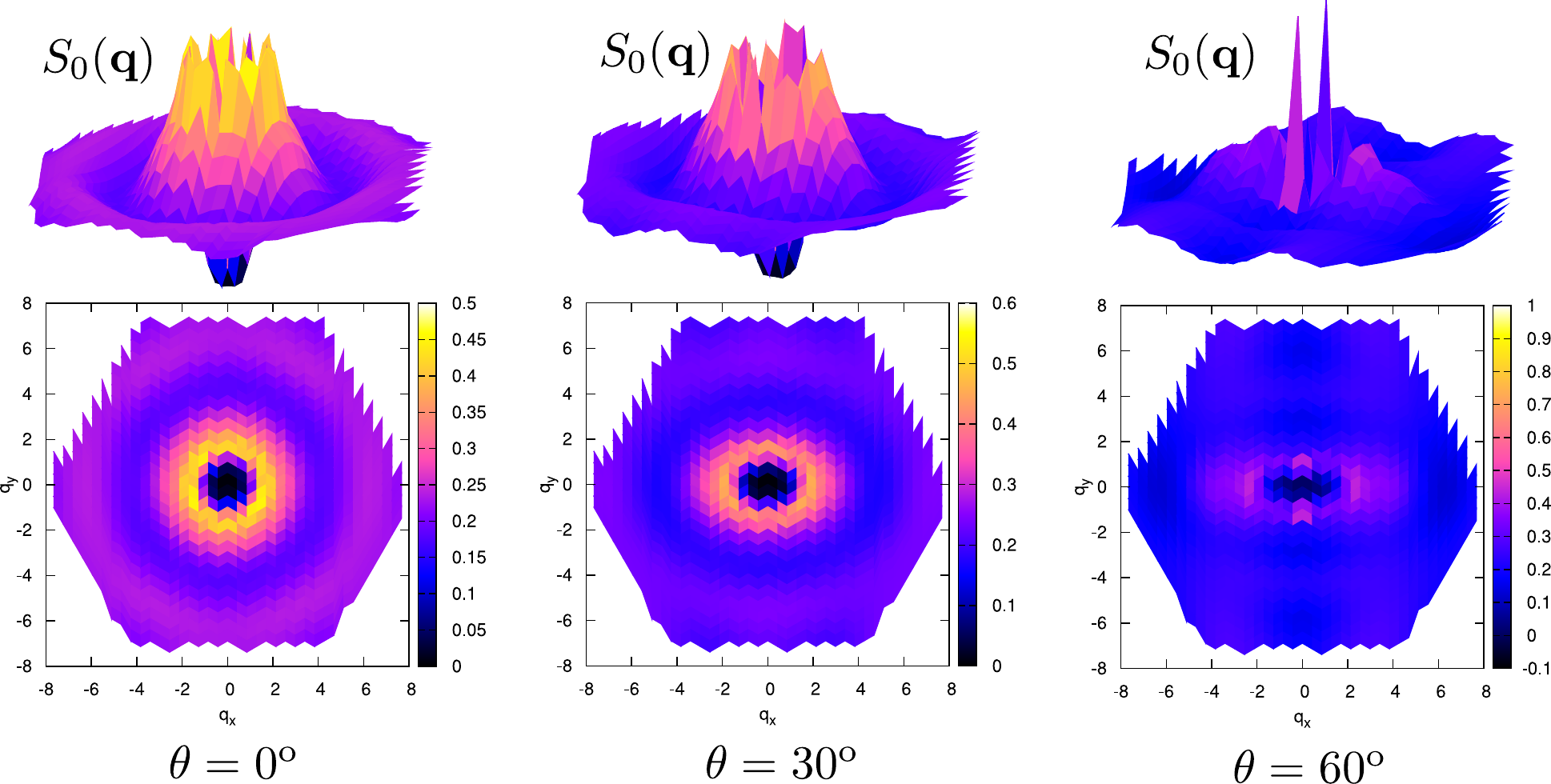}
  \end{minipage}
\caption{(color online) Guiding center structure factor for $N=10$ particles at $\nu=1/3$ filling and tilt angle zero, 30 and 60 degrees (left to right). The state at the largest tilt angle has charge-density order reflected in two sharp peaks, while for two smaller tilt angles the ground state is a liquid.}
\label{fig_laughlinsq}
\vspace{-0pt}
\end{figure*}

To evaluate the matrix element, it is convenient to use the Fourier transform
$$
V(\mathbf{r}_1-\mathbf{r}_2,z_1-z_2)=\sum_{\mathbf{q}}\int dq_z \frac{1}{\mathbf{q}^2+q_z^2}e^{i\mathbf{q}(\mathbf{r}_1-\mathbf{r}_2)}e^{i q_z (z_1-z_2)}.
$$
Note that this is a Fourier transform of the 3D Coulomb potential, but at end we will integrate out $q_z$.
Finally, the torus matrix element for subband (0,0) reads
\begin{eqnarray}\label{matel}
\nonumber V_{j_1...j_4}&=&\frac{\delta_{j_1+j_2,j_3+j_4}'}{N_{\phi}}\sum_{q_x=\frac{2\pi s}{a},q_y=\frac{2\pi t}{b}}' \delta_{q_y,X_{j_1}-X_{j_4}}' e^{-i q_x (X_{j_1}-X_{j_3})} \\ 
\nonumber && e^{-\frac{q_x^2}{2} (\ell_1^2 \cos^2\phi + \ell_2^2 \sin^2\phi)-\frac{q_y^2}{2} \left( \frac{\cos^2\phi}{\ell_1^2} + \frac{\sin^2\phi}{\ell_2^2} \right) } \\
\nonumber && \int dq_z \frac{1}{q_x^2+q_y^2+q_z^2} e^{-\frac{1}{2}q_z^2 (\ell_2^2\cos^2\phi + \ell_1^2\sin^2\phi) } \\
&& e^{ q_z q_x \sin\phi\cos\phi (\ell_1^2-\ell_2^2) }
\end{eqnarray}
The prime on the delta functions stands for ``modulo $N_\phi$", and the prime on the summation indicates that $\mathbf{q}=0$ component has been cancelled out by the neutralizing (positive) backround charge. In the case of (0,0) subband, the above matrix element can be analytically further simplified to some extent, but this is no longer the case when higher subbands are considered, and one is left with the general expression quoted in Eq.~(\ref{matel}). 

The effective matrix element projected to one of the higher subbands is obtained by multiplying the integrand in Eq.~(\ref{matel}) by an extra form factor $|F(\mathbf{q},q_z)|^2$. The computation of this form factor is straightforward and involves the standard algebra of Landau level raising/lowering operators~\cite{prange}, but it quickly becomes tedious for very high subbands. Here we quote the result for the excited subband (0,1) where $F(\mathbf{q},q_z)$ is given by
$$
F^{01} (\mathbf{q},q_z) = 1 - \frac{1}{2\ell_2^2}q_y^2 \sin^2\phi - \frac{1}{2}\ell_2^2 (q_x\sin\phi + q_z \cos\phi)^2,
$$
and for (1,0) subband
\begin{eqnarray}
\nonumber F^{10} (\mathbf{q},q_z)=1 &-& \frac{1}{2}(\ell_1^2 q_x^2 + q_y^2/\ell_1^2) \cos^2\phi \\
 \nonumber &-& \frac{1}{2}\ell_1^2 q_z \sin\phi (-2 q_x\cos\phi + q_z \sin\phi).
\end{eqnarray}
Altogether, the matrix elements have a somewhat complicated form, but all the intermediate integrals/sums converge rapidly, hence can be straightforwardly evaluated in practice. Therefore, one can follow the standard approach of diagonalizing the many-body Hamiltonian in a basis of periodic orbitals on the surface of the torus~\cite{duncan_translations}. To reduce the computational cost, it is desirable to use invariance under magnetic translations to block-reduce the Hamiltonian. This formalism was first given in Ref.~\onlinecite{duncan_translations} (see Refs.~\onlinecite{chakpiet, andrei_translations} for pedagogical reviews).

Note that from the form of the Hamilton (\ref{matel}) it is clear that the parallel field explictly builds in  anisotropy in the problem: the components $q_x$ and $q_y$ in the Gaussian factor are coupled to different effective magnetic lengths. This can be understood semiclassically: in a purely perpendicular field, the cyclotron orbits of an electron are circles. When the field is tilted, electrons orbit around the tilted axis. Due to the confinement of electrons to the 2D layer, the true shape of their orbits is a projection of these circles to the 2D plane, i.e. their shape becomes elliptical.  

\begin{figure}[htb]
\includegraphics[scale=0.4,angle=0]{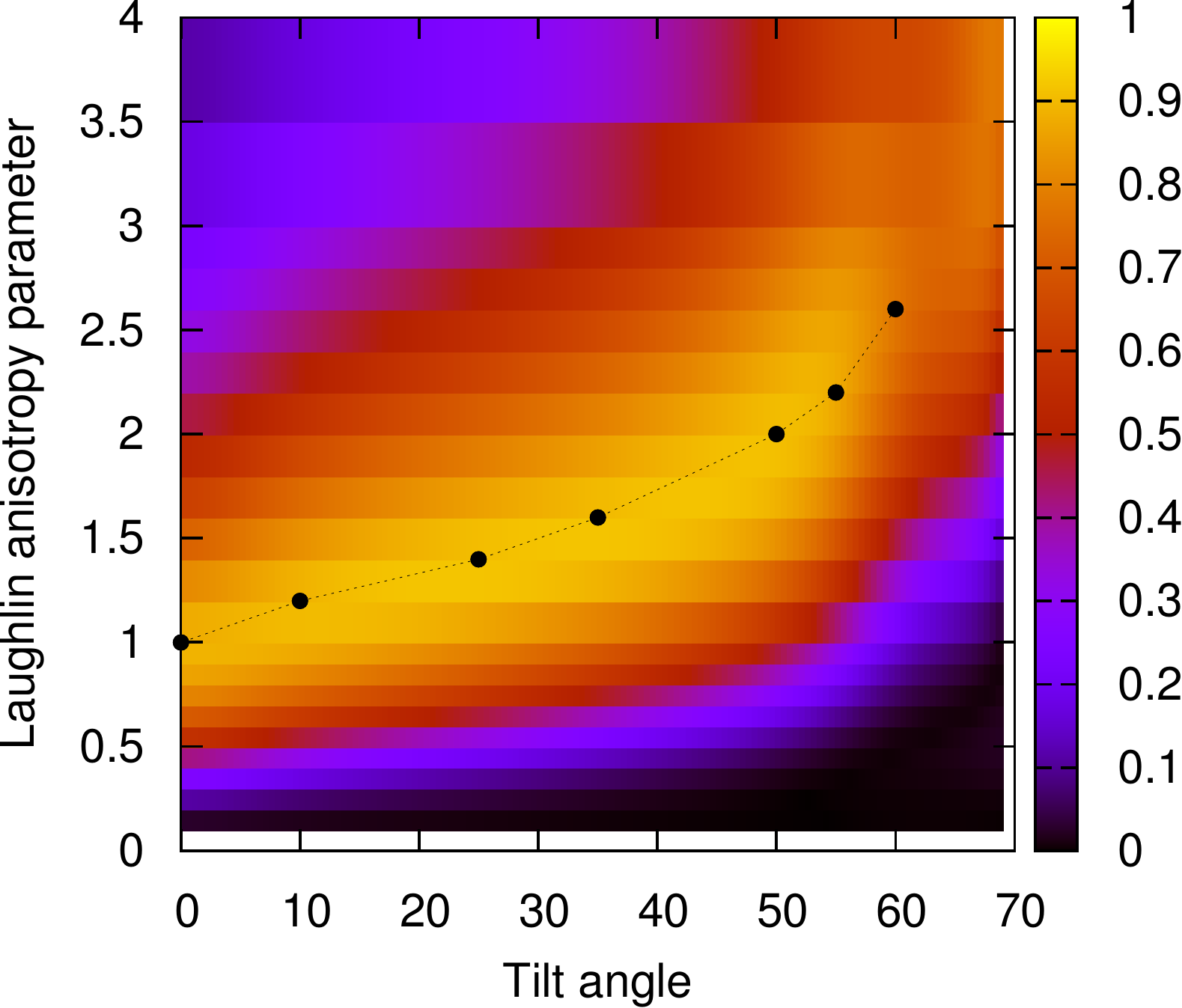}
\caption{(color online) Overlap between the exact ground state at each tilt angle $\theta$, $\Psi_0(\theta)$, and a family of Laughlin states parametrized by the metric of their fundamental droplet, $\Psi_L^g$. Color scale represents the value of the overlap as a function of tilt angle and parameter $\alpha$ that defines the metric. Black points indicate the maximal overlap for the given tilt angle.}
\label{fig_metric}
\end{figure}

\section{Results}\label{sec_results}

\begin{figure*}[ttt]
  \begin{minipage}[l]{\linewidth}
    \includegraphics[scale=0.6]{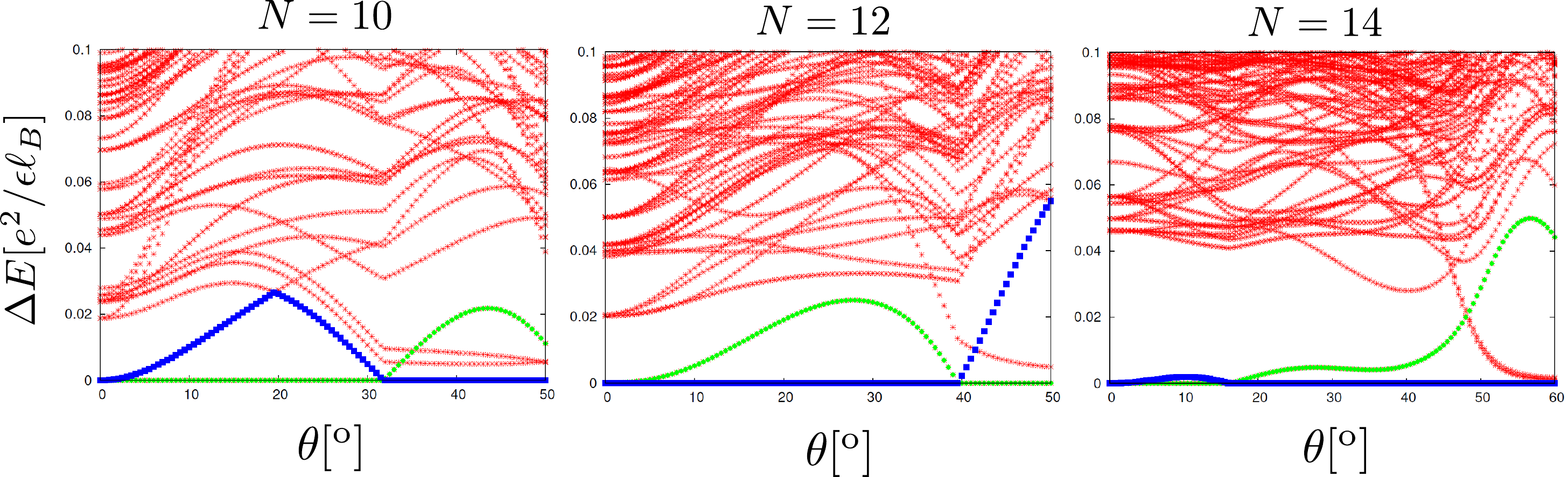}
  \end{minipage}
\caption{(color online) Energy spectrum for $N=10$, 12,and 14 particles at filling 1/2 in the subband (1,0) as a function of tilt angle (measured in degrees). Spectrum is plotted relative to the ground state at each tilt angle, illustrating the behavior of the neutral gap of the system. Levels tentatively belonging to the topologically degenerate Moore-Read ground states are denoted by blue and green colors.}
\label{fig_onehalf}
\vspace{-0pt}
\end{figure*}
In this Section we present the results of exact diagonalization of the Hamiltonian (\ref{matel}) for a finite number of electrons in a toroidal geometry. This choice of boundary condition is dictated by two requirements: (i) compatibility of the Landau gauge with the presence of a parallel field, and (ii) the absence of open boundaries which avoids the complications due to the edge physics. The presence of parallel field does not affect the standard symmetry classification, as we mentioned in Sec.~\ref{sec_manybody}, Haldane's many-body momentum~\cite{duncan_translations} can be used to label the states. We will consider hexagonal unit cells, and focus on partial fillings 1/3 and 1/2 of the subband (1,0). This corresponds to filling factors $\nu=7/3$ and $\nu=5/2$ which have been the subject of recent experimental studies~\cite{xia2011,tsui_52,shayegan_tilt}. Also, for simplicity we fix $\omega_0/\omega_z=1.3$, which roughly corresponds to the sample design in Ref.~\onlinecite{shayegan_tilt}. (The main conclusions do not depend on the precise value of this ratio.) Note that in the simulations presented here we neglect the spin of the electrons, and only concentrate on the orbital effects of the tilt because the ground states at $\nu=7/3$ and $\nu=5/2$ are believed to be polarized~\cite{morf_pf,52_spin}.

In Fig.~\ref{fig_onethird} we show the energy spectrum of $N=10$ particles at filling 1/3 of the subband (1,0) as a function of tilt angle (measured in degrees). Inset shows the same spectrum but plotted relative to the ground state at each tilt angle. In other words, the inset illustrates the behavior of the neutral gap of the system as a function of tilt (in units of $e^2/\epsilon \ell_0$). Note that the experimentally measured gaps in transport correspond to the so-called ``charge" gaps, which are significantly harder to compute in the periodic geometry (for the purpose of computing charge gaps, sphere geometry has been used, almost universally, in the literature~\cite{jainbook}).

The evolution depicted in Fig.~\ref{fig_onethird} suggests that $\nu=7/3$ is fairly robust for small tilt angles (up to 30-40 degrees). The neutral gap even appears to increase for small tilt angles up to 10 degrees, however a careful extrapolation of the gap as a function of $1/N$ would be needed to firmly conclude whether the state becomes enhanced for small tilt as some of the experiments seemed to indicate~\cite{tsui_52}. Beyond 40 degrees, the ground state energy sharply rises, and states from different $\mathbf{k}$ sectors join to form a quasidegenerate ground-state manifold. Upon closer examination, we find that the momenta of the quasidegenerate states are of the form $(k_x,0)$, indicating linear order along $x$-direction. This is known to be one of the signatures of the stripe phase~\cite{anisotropic,kfs_prl,kfs_prb,moessner_chalker,edduncanky_stripe}.

\begin{figure}[htb]
\includegraphics[scale=0.4,angle=0]{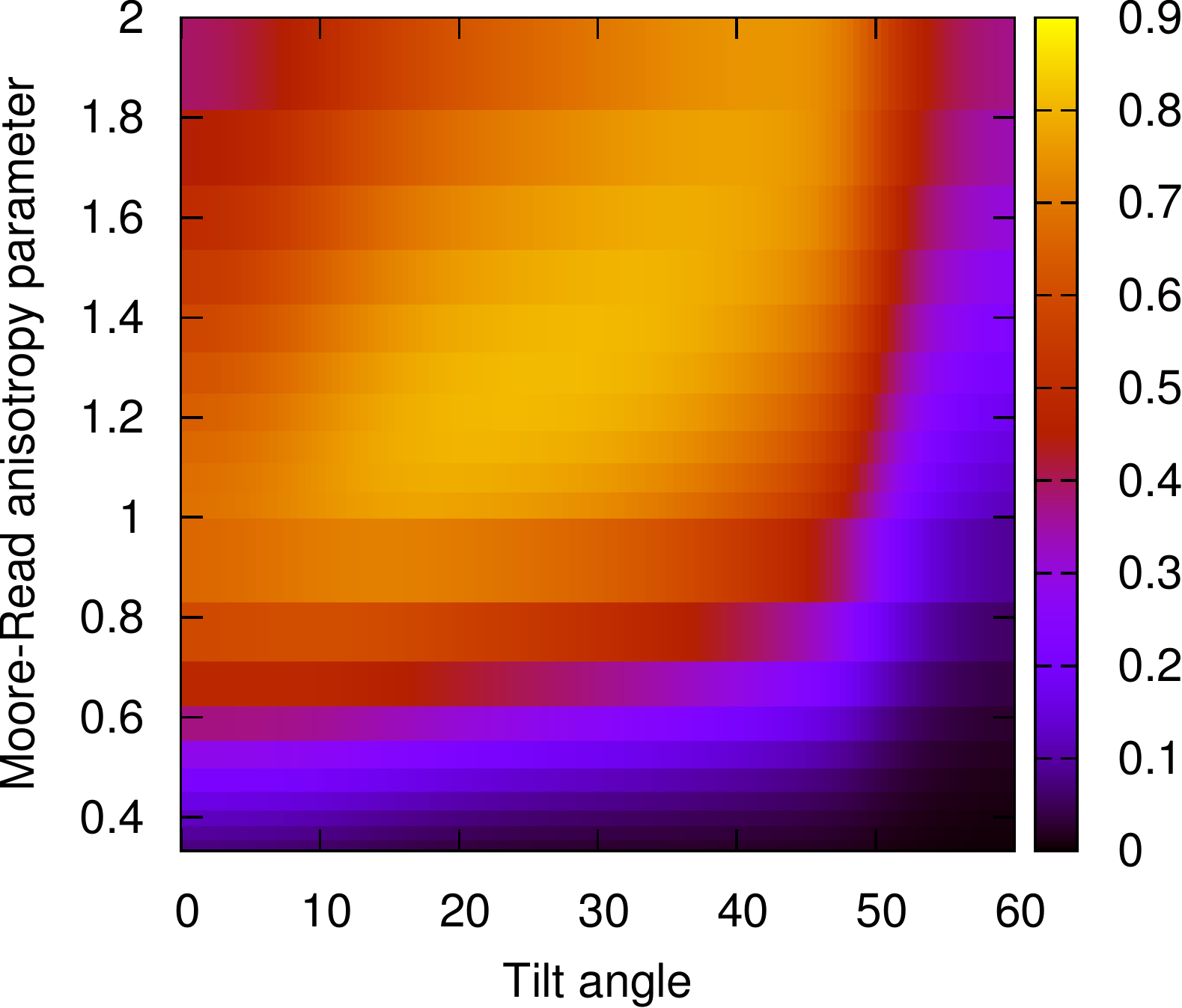}
\caption{(color online) Overlap between the exact ground state at each tilt angle $\theta$, $\Psi_0(\theta)$, and a family of Moore-Read states parametrized by the metric of their fundamental droplet, $\Psi_{\rm MR}^g$. Color scale represents the value of the overlap as a function of the tilt angle and the parameter $\alpha$ that defines the metric. System size is $N=14$ particles on a hexagonal torus in (0,7) momentum sector.}
\label{fig_metric2}
\end{figure}
Further evidence for the stripe phase is found in the shape of the guiding center structure factor~\cite{haldane_prange},
\begin{equation}\label{sq}
S_0(\mathbf{q}) = \frac{1}{N_\phi} \sum_{i,j}\langle e^{i\mathbf{q}\cdot \mathbf{R}_i} e^{-i\mathbf{q}\cdot \mathbf{R}_j} \rangle - \langle e^{i\mathbf{q}\cdot \mathbf{R}_i} \rangle \langle e^{-i\mathbf{q}\cdot \mathbf{R}_j} \rangle,
\end{equation}
as a function of tilt, as shown in Fig.~\ref{fig_laughlinsq}. In the definition of $S_0(\mathbf{q})$, the brackets $\langle ... \rangle$ denote the ground-state expectation value of the operator representing the Fourier component of the guiding-center density, 
\begin{eqnarray}\label{density}
\rho (\mathbf{q}) = \sum_i^N e^{i\mathbf{q}\cdot \mathbf{R}_i}.
\end{eqnarray}
Here $\mathbf{R}_i$ denotes the guiding center coordinate~\cite{prange} of the particle $i$. Compared to the standard definition~\cite{sma}, $S_0$ has the single-particle form-factors stipped off, and it is normalized per flux quantum instead of per particle. For zero tilt, $S_0$ has a characteristic circular maximum, and tends to a constant value $\nu-\nu^2$ (for the normalization chosen above) as $q \to \infty$~\cite{haldane_prange}. For tilt angle of 30 degrees, the structure factor has a similar behavior as a function of $q_x,q_y$, but the locus of its maxima has become an ellipse stretched along $x$-direction. Finally, for tilt of 60 degrees, $S_0$ displays a qualitatively different shape with two sharp peaks, indicating broken-symmetry ordering in the $x$-direction. This is another evidence~\cite{edduncanky_stripe} for the formation of the stripe phase in the regime of tilt beyond 40 degrees.  

\begin{figure*}[ttt]
  \begin{minipage}[l]{\linewidth}
    \includegraphics[scale=0.8]{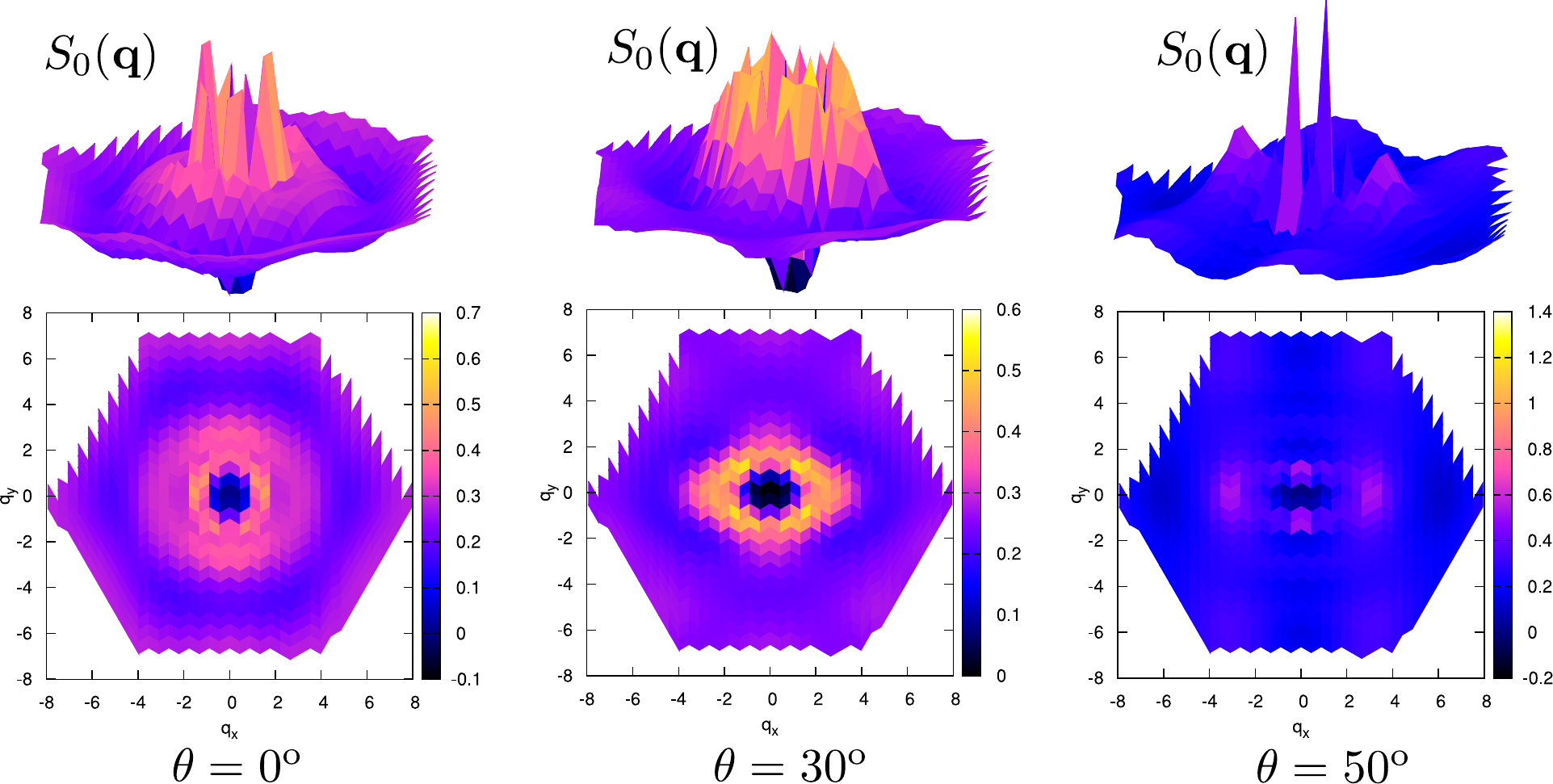}
  \end{minipage}
\caption{(color online) Guiding center structure factor for a particular member of the ground state manifold (labeled $(0,7)$ in Fig.~\ref{fig_onehalf}) for $N=14$ particles at $\nu=1/2$ in (1,0) subband, for tilt angle 0, 30 and 50 degrees. The state at the largest tilt angle has charge-density order reflected in two sharp peaks, whereas for smaller tilt angles the ground state is a liquid.}
\label{fig_mrsq}
\vspace{-0pt}
\end{figure*}

In the regime between zero and 30 degree tilt, we identify the ground state as a FQH liquid corresponding to the family of Laughlin wavefunctions with the internal metric fluctuating to optimize itself with respect to the external perturbation (tilt). The existence of this internal degree of freedom was recently pointed out in Ref.~\onlinecite{duncan_unpublished}. A convenient way of formally defining the family of FQH liquid states with a varying internal metric is through the pseudopotential formalism~\cite{haldane_prange}. In the case of the Laughlin 1/3 filling, this family of states $\{ \Psi_L^g \}$ is defined by the zero-mode condition
\begin{eqnarray}
\hat V_1 (g) \Psi_L^g = 0,
\end{eqnarray}  
where $\hat V_1(g)$ is the generalized Haldane pseudopotential Hamiltonian with a given metric $g$~\cite{duncan_unpublished}:
\begin{eqnarray}\label{v1}
\hat V_1 (g) = \sum_{\mathbf{q}} L_1(q_g^2\ell_B^2) \exp(-\frac{q_g^2\ell_B^2}{2}) \rho (\mathbf{q}) \rho (-\mathbf{q}),
\end{eqnarray} 
Here the operator $\rho$ is defined in Eq.(\ref{density}), and $L_1$ is the first Laguerre polynomial.
Two-dimensional metric $g$ defines the norm of $q_g^2 \equiv g^{ab}q_a q_b$ and is taken to be unimodular $\det g =1$. The first-quantized expressions for $\Psi_L^g$ are given in Ref.~\onlinecite{kunyang_aniso}.

Previously, generalized Laughlin wavefunctions have been studied in the context of anisotropic systems~\cite{boyang,xinwan}, where it was assumed that the band mass tensor or the Coulomb dielectric tensor is explicitly anisotropic. These works have established that FQHE physics survives some amount of anisotropy, while the phase corresponding to large anisotropy was identified with a stripe. In Ref.~\onlinecite{xinwan} some evidence for a quantum Hall nematic phase~\cite{nematic} was provided for the intermediate regime of anisotropies. As we mentioned in Sec.~\ref{sec_manybody}, the dominant Gaussian term in the tilted Hamiltonian has the same form as in mass-anisotropic systems, therefore we might expect some similarity with the results obtained in Refs.~\onlinecite{boyang,xinwan}. However, the presence of extra terms in Eq.~(\ref{matel}) prevents a complete mapping of the tilted field onto a mass-anisotropic problem.

To further motivate the identification of the small-tilt phase with the FQH liquid, we compute the overlap between the exact ground state at a given tilt, and the family of Laughlin states $\Psi_L^g$ parametrized the anisotropy parameter $\alpha$, Fig.~\ref{fig_metric}. We assume $g$ is unimodular, and does not contain any off-diagonal terms. Therefore, the metric $g$ is parametrized by $\alpha,1/\alpha$ on the diagonal. By varying $\alpha$, for each tilt angle we find it is possible to achieve a high overlap (in excess of 97\%), typical of the Laughlin state. The maximum overlap can be used as a criterion for determining the internal metric of the Laughlin state at a given tilt. 
As seen in Fig.~\ref{fig_metric}, the maximum overlap varies approximately linearly with tilt for tilt angles smaller than 40 degrees, in agreement with the result found in systems with band anisotropy~\cite{boyang}.
 
Another fractional state of interest in (1,0) subband is the $\nu=1/2$, corresponding to the total filling $\nu=5/2$ in experiment. In GaAs heterostructure samples without tilt, this state is roughly of the same strength as $\nu=7/3$~\cite{125}. A great deal of theoretical evidence~\cite{morf_pf,rh_pf} points to the fact that $\nu=5/2$ state is described by the Moore-Read Pfaffian wavefunction, and might have non-Abelian quasiparticles in the bulk, which makes it more exotic than the states of the Laughlin type. A necessary requirement for the non-Abelian statistics in this case is the full spin polarization of the ground state, which is consistent with theoretical predictions~\cite{morf_pf} as well as experimental findings~\cite{52_spin}. 

In Fig.~\ref{fig_onehalf} we show the energy spectrum at filling 1/2 of the subband (1,0) as a function of tilting angle (in degrees). This state is more fragile than the 1/3 state, therefore we show three different system sizes ($N=10,12,14$) to illustrate the convergence to the thermodynamic limit. As before, the data in the upper panel in Fig.~\ref{fig_onehalf} corresponds to the raw energy spectrum, and the lower panel shows the neutral gap.  

The difficulty in establishing convergence to the thermodynamic limit in this case is partly related to the fact that the Moore-Read state has a 6-fold topological degeneracy on the torus. Using the conventional symmetry classification, this degeneracy can be factored into a trivial 2-fold degeneracy~\cite{duncan_translations}, and the residual 3-fold degeneracy resulting from the non-Abelian statistics~\cite{readgreen}. The 3-fold degenerate states belong to the same momentum sector only in case of the hexagonal symmetry. Tilted field, however, reduces the symmetry down to centered rectangular, and only two of the states remain in the same sector. In finite systems, there will be some amount of splitting between the sectors, as can be seen in Fig.~\ref{fig_onehalf}. As the $N=14$ system shows, the splitting gets suppressed as larger sizes are approached and is consistent with eventually becoming zero in thermodynamic limit. The fact that there is a well-defined 3-fold ground state multiplet even for the small systems that can be accessed by exact diagonalization is an important piece of evidence in favor of the Moore-Read state.    

Apart from topological degeneracy, it is also possible to compute the overlap between the exact ground state and the Moore-Read wavefunction. Similar to the Laughlin case discussed above, in a situation where tilt is present, one must consider a family of Moore-Read states parametrized by the internal metric. The overlaps between this family of states and the exact ground state are given in Fig.~\ref{fig_metric2}. The behavior of the maximum overlap as a function of tilt is qualitatively consistent with Fig.~\ref{fig_metric}. The value of the maximum overlap with the Moore-Read state is smaller than in case of the Laughlin state, but it can improved by varying some of the short-range Haldane pseudopotentials, as it has been done in the literature~\cite{rh_pf}.

Apart from some amount of splitting within the ground state manifold, the evolution of the energy spectrum in Fig.~\ref{fig_onehalf} is also reminiscent of the 1/3 case (Fig.~\ref{fig_onethird}). The quantum Hall state remains stable up to 30-40 degree tilt, at which point it gives rise to a stripe phase. This transition is again captured in the behavior of the structure factor displayed in Fig.~\ref{fig_mrsq}.

\section{Discussion and conclusions}\label{sec_conc}

In this paper we have studied the effect of coupling between a parallel component of the magnetic field and the electronic motion restricted to the 2D plane of a parabolic quantum well. We have demonstrated that this coupling probes the geometrical degree of freedom of fractional quantum Hall liquid states~\cite{duncan_unpublished}. For filling factors 1/3 and 1/2 of the excited (1,0) subband, the ground states are  incompressible fluids when the parallel field is zero. As the tilt angle is increased from zero to 30 degrees, the internal geometry of these states adjusts itself to accommodate the variation of the external metric imposed by the tilt. Up to this distortion of the elementary droplets from circular to elliptical, the FQHE physics is maintained in this regime. Beyond 40 degrees tilt, the states undergo a transition to the broken-symmetry phase with stripe order. 

\begin{figure}[htb]
\includegraphics[scale=0.42,angle=0]{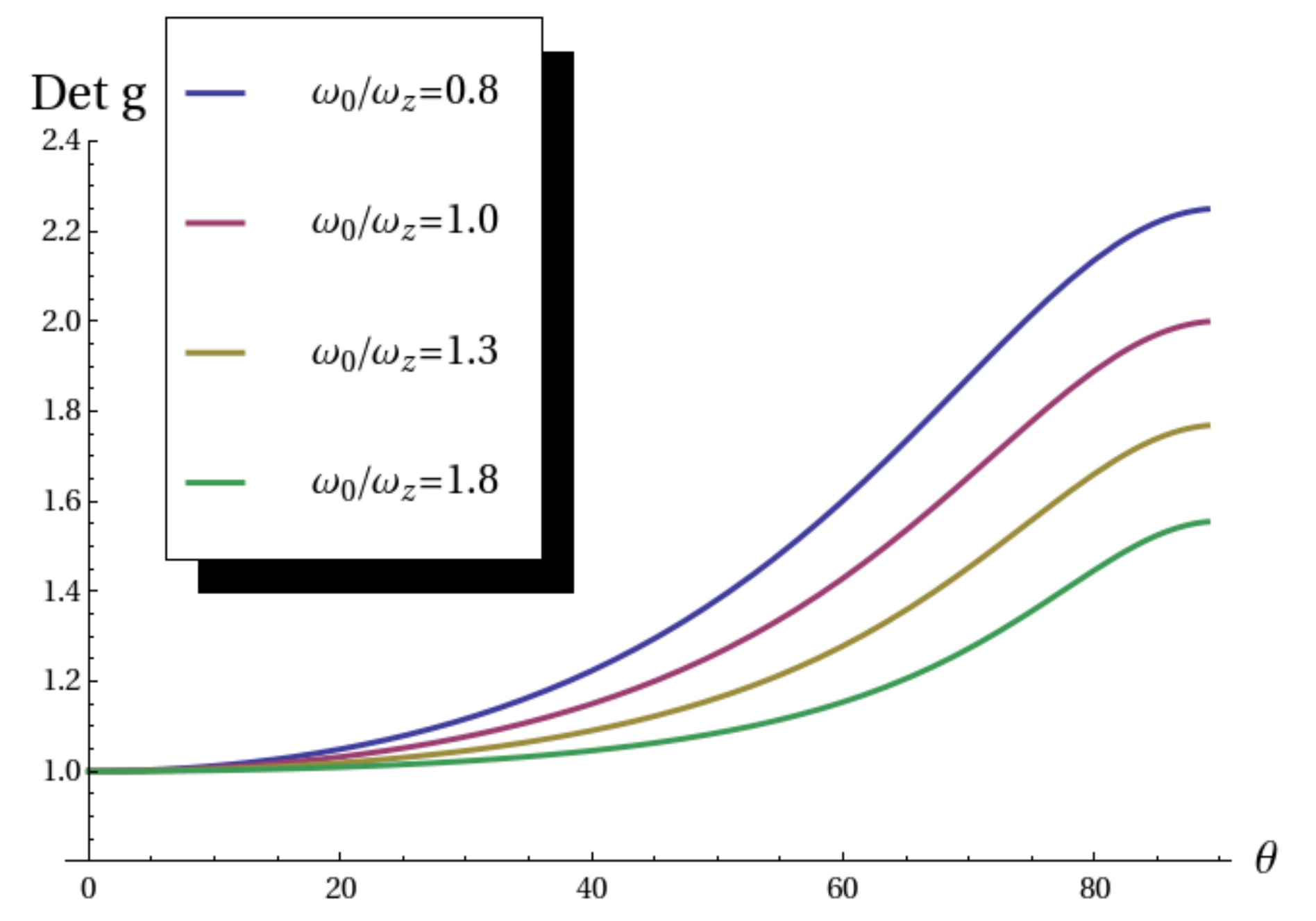}
\caption{(color online) Determinant of the metric in the Gaussian form-factor of Eq.~(\ref{matel}) as a function of tilt, for several values of the confining potential.}
\label{fig_detg}
\end{figure}

Previously, FQHE was studied in systems with explicit anisotropy introduced through the band-mass tensor~\cite{boyang} or the dielectric tensor defining the Coulomb interaction~\cite{xinwan} (see also experimental results for the anisotropic ``composite Fermi liquid" state in Ref.~\onlinecite{kamburov}). For small tilt angles, our results are in qualitative agreement because the Gaussian factor (\ref{matel}) contains a metric $g$ that is nearly unimodular, and hence (neglecting the complicated prefactors) the tilted problem can be rewritten as an effective mass anisotropy. The ``volume" is given by $\det g= (\ell_1^2 \cos^2\phi + \ell_2^2 \sin^2\phi)( \cos^2\phi/\ell_1^2 + \sin^2\phi/\ell_2^2 )$, and $\det g$ is plotted as a function of tilt in Fig.~\ref{fig_detg}. The volume remains fairly close to unity for tilts smaller than 40 degrees, supporting the similarity between tilt and mass-anisotropy. Note that lower values of the confinement $\omega_0/\omega_z$ lead to a faster deviation of the volume away from unity, and therefore the intrinsic tilt effects are effectively stronger in this regime. Furthermore, it would be interesting to explore connections between a spatially non-uniform tilt and the so-called Fubini-Study metric in lattice analogs of FQHE~\cite{fubini}.

In principle, the tilted-field phase diagram might also admit the existence of the so-called nematic quantum Hall phases~\cite{nematic}. These phases are believed to spontaneously break rotational symmetry, whereas the tilted field breaks such a symmetry explicitly. Although we do not find direct evidence for such phases in the case of generic (Coulomb) interaction studied here, it would be important to understand better the connection between nematic phases~\cite{mulligan,maciejko} and quantum Hall phases with elliptical elementary droplets, and how one could tune between them by modifying the interaction.    

Finally, we mention several limitations of the present work. Apart from the ``obvious" assumptions of zero temperature and clean (translationally invariant) systems, the main limitations of the present work are the neglect of the electron spin, and the mixing between various subbands. Incorporating spin would be useful for other fractions, notably $\nu=2/5$ (corresponding to $\nu=12/5$ in experiment) where a spin transition was detected as a function of tilt~\cite{tsui_125}. However, this fraction has a much smaller gap in comparison to 1/3 or 1/2, and numerical techniques beyond exact diagonalization are required to access sufficiently large systems that are not plagued by the finite-size effects. Similarly, the mixing between different subbands is another effect that is difficult to treat reliably within exact-diagonalization schemes. While we do not expect our results to be strongly affected by the mixing when tilt is less than 40-50 degrees, in the regime of large tilts ($\geq 70$ degrees) they will undoubtedly be important, as we can see from the Landau level structure (Fig.~\ref{fig_tilt}). With respect to the results presented here, mixing effects are furthermore important for two specific reasons. Firstly, the mixing lifts the particle-hole symmetry between the Pfaffian and anti-Pfaffian~\cite{llmix}, and selects only one of them to describe the ground state at $\nu=5/2$. Secondly, experiments performed in the regime of extremely large tilts ($\sim$70 degrees) have found a puzzling re-emergence of \emph{isotropic} transport~\cite{xia2011} in that limit. It would be interesting to understand the connection between isotropic transport and extreme LL mixing in future work.  

\section{Acknowledgments}

This work was supported by DOE grant DE-SC$0002140$. I would like to thank S. Johri, R. Bhatt, J. Maciejko, E. Rezayi and M. Milovanovi\'c for useful comments, and Y. Liu and M. Shayegan for discussing experimental data.


\begin{thebibliography}{9}

\bibitem{tsg}
D. C. Tsui, H. L. Stormer, and A. C. Gossard, Phys. Rev. Lett. {\bf 48}, 1559 (1982).

\bibitem{prange}
\emph{The Quantum Hall Effect}, 2nd ed., edited by R. E. Prange and S. M. Girvin, Springer-Verlag, New York, 1990.

\bibitem{laughlin}
R. B. Laughlin, Phys. Rev. Lett. {\bf 50}, 1395 (1983).

\bibitem{jack} B. Andrei Bernevig and F. D. M. Haldane, Phys. Rev. Lett. {\bf 100}, 246802 (2008).

\bibitem{zhk} S. C. Zhang, T. H. Hansson, and S. Kivelson, Phys. Rev. Lett. {\bf 62}, 82 (1989).

\bibitem{mr}
G. Moore and N. Read, Nucl. Phys. B {\bf 360}, 362 (1991).

\bibitem{jainbook} J. K. Jain, \emph{Composite fermions}, (Cambridge University Press, 2007).

\bibitem{ed} F. D. M. Haldane and E. H. Rezayi, Phys. Rev. Lett. {\bf 54}, 237 (1985).

\bibitem{haldane_sphere} F. D. M. Haldane, Phys. Rev. Lett. {\bf 51}, 605 (1983).

\bibitem{haldane_prange} F. D. M. Haldane in Ref.~\onlinecite{prange}.

\bibitem{halperin} B. I. Halperin, Phys. Rev. Lett. {\bf 52}, 1583, 2390(E) (1984).

\bibitem{fqhe_graphene} X. Du \emph{et al.}, Nature {\bf 462}, 192 (2009); K. Bolotin \emph{et al.}, Nature {\bf 462}, 196 (2009); C. R. Dean \emph{et al.}, Nat. Phys. {\bf 7}, 693 (2011); F. Ghahari \emph{et al.}, Phys. Rev. Lett. {\bf 106}, 046801 (2011); B. E. Feldman \emph{et al.}, Science {\bf 337}, 1196 (2012).

\bibitem{graphene_multicomponent} Kun Yang, S. Das Sarma, and A. MacDonald, Phys. Rev. B {\bf 74}, 075423 (2006); C. T\H{o}ke and J. K. Jain, Phys. Rev. B {\bf 75}, 245440 (2007); Z. Papi\'c, M. O. Goerbig, and N. Regnault, Phys. Rev. Lett. {\bf 105}, 176802 (2010); Z. Papi\'c, D. A. Abanin, Y. Barlas, and R. N. Bhatt, Phys. Rev. B {\bf 84}, 241306  (2011). 

\bibitem{llmix} Waheb Bishara and Chetan Nayak, Phys. Rev. B {\bf 80}, 121302 (2009);
Arkadiusz W\'ojs, Csaba T\H{o}ke, and Jainendra K. Jain, Phys. Rev. Lett. {\bf 105}, 096802 (2010); 
Edward H. Rezayi and Steven H. Simon, Phys. Rev. Lett. {\bf 106}, 116801 (2011); I. Sodemann and A. MacDonald,  arXiv:1302.3896 (2013); M. Peterson and C. Nayak,  arXiv:1303.1541 (2013); S. H. Simon and E. H. Rezayi, Phys. Rev. B {\bf 87}, 155426 (2013).



\bibitem{52}  R. Willett, \emph{et al.}, Phys. Rev. Lett. {\bf 59}, 1776 (1987).

\bibitem{125}  J.S. Xia, \emph{et al.}, Phys. Rev. Lett. {\bf 93}, 176809 (2004); W. Pan, \emph{et al.}, Phys. Rev. B {\bf 77}, 075307 (2008); A. Kumar \emph{et al.}, Phys. Rev. Lett. {\bf 105}, 246808 (2010). 

\bibitem{rr}  N. Read, and E. Rezayi, Phys. Rev. B. {\bf 59}, 8084 (1999)

\bibitem{llmix_shayegan} Yang Liu, D. Kamburov, M. Shayegan, L.N. Pfeiffer, K.W. West, K.W. Baldwin, Phys. Rev. Lett {\bf 107}, 176805 (2011). 

\bibitem{llmix_csathy} N. Samkharadze, L.N. Pfeiffer, K.W. West, G.A. Cs\'athy, arXiv:1302.1444 (2013).


\bibitem{antipf} M. Levin, B. I. Halperin, and B. Rosenow, Phys. Rev. Lett. {\bf 99}, 236806 (2007); S.-S. Lee, S. Ryu, C. Nayak, and M. P. A. Fisher, Phys. Rev. Lett. {\bf 99}, 236807 (2007).

\bibitem{zed} Z. Papi\'c, F. D. M. Haldane, and E. Rezayi, Phys. Rev. Lett. {\bf 109}, 266806 (2012).


\bibitem{tilt_old} J. P. Eisenstein, R. Willett, H. L. Stormer, D. C. Tsui, A. C. Gossard, and J. H. English
Phys. Rev. Lett. {\bf 61}, 997 (1988).

\bibitem{engel} L. W. Engel, S. W. Hwang, T. Sajoto, D. C. Tsui, and M. Shayegan, Phys. Rev. B {\bf 45}, 3418 (1992).

\bibitem{csathy_tilt} G. A. Cs\'athy, J. S. Xia, C. L. Vicente, E. D. Adams, N. S. Sullivan, H. L. Stormer, D. C. Tsui, L. N. Pfeiffer, and K. W. West, Phys. Rev. Lett. {\bf 94}, 146801 (2005). 

\bibitem{dean} C. R. Dean, B. A. Piot, P. Hayden, S. Das Sarma, G. Gervais, L. N. Pfeiffer, and K. W. West, Phys. Rev. Lett. {\bf 101}, 186806 (2008).

\bibitem{du} Chi Zhang, T. Knuuttila, Yanhua Dai, R. R. Du, L. N. Pfeiffer, and K. W. West, Phys. Rev. Lett. {\bf 104}, 166801 (2010). 

\bibitem{xia_tilt} J. Xia, V. Cvicek, J. P. Eisenstein, L. N. Pfeiffer, and K. W. West, Phys. Rev. Lett. {\bf 105}, 176807 (2010). 

\bibitem{xia2011} J. Xia, J. P. Eisenstein, L. N. Pfeiffer, and K. W. West, Nature Phys. {\bf 7}, 845 (2011).

\bibitem{tsui_52} Guangtong Liu, Chi Zhang, D. C. Tsui, Ivan Knez, Aaron Levine, R. R. Du, L. N. Pfeiffer, and K. W. West, Phys. Rev. Lett. {\bf 108}, 196805 (2012).

\bibitem{tsui_125} Chi Zhang, Chao Huan, J. S. Xia, N. S. Sullivan, W. Pan, K. W. Baldwin, K. W. West, L. N. Pfeiffer, and D. C. Tsui, Phys. Rev. B {\bf 85}, 241302 (2012).

\bibitem{shayegan_tilt} Yang Liu, S. Hasdemir, M. Shayegan, L.N. Pfeiffer, K.W. West, K.W. Baldwin,  arXiv:1302.6304 (2013).


\bibitem{maan} J. C. Maan, in \emph{Two Dimensional Systems, Heterostructures and Superlattices}, edited by G. Bauer, F. Kucher, and H, Heinrich (Springer, Heidelberg, 1984).

\bibitem{chak} V. Halonen, P. Pietilainen and T. Chakraborty, Phys. Rev. B {\bf 41}, 10202 (1990).

\bibitem{demler} Daw-Wei Wang, Eugene Demler, and S. Das Sarma, Phys. Rev. B {\bf 68}, 165303 (2003).

\bibitem{peterson} Michael R. Peterson, Th. Jolicoeur, and S. Das Sarma, Phys. Rev. B {\bf 78}, 155308 (2008).  

\bibitem{yhl} D. Yoshioka, B. I. Halperin, and P. A. Lee, Phys. Rev. Lett. {\bf 50}, 1219 (1983).

\bibitem{duncan_translations} F. D. M. Haldane, Phys. Rev. Lett. {\bf 55}, 2095 (1985).

\bibitem{chakpiet}  T. Chakraborty and P. Pietilainen, \emph{The Quantum Hall
Effects}, Springer, New York (1995).

\bibitem{andrei_translations} B. A. Bernevig and N. Regnault, Phys. Rev. B {\bf 85}, 075128 (2012). 

\bibitem{morf_pf}
R. H. Morf, Phys. Rev. Lett. {\bf 80}, 1505 (1998).

\bibitem{52_spin} M. Stern \emph{et al.}, arXiv:1201.3488 (2012); L. Tiemann \emph{et al.}, arXiv:1201.3737 (2012).
X. Lin \emph{et al.}, arXiv:1201.3648 (2012).


\bibitem{anisotropic} M. P. Lilly, K. B. Cooper, J. P. Eisenstein, L. N. Pfeiffer, and K. W. West, Phys. Rev. Lett. {\bf 82}, 394 (1999); 
R. R. Du, D. C. Tsui, H. L. Stormer, L. N. Pfeiffer, K. W. Baldwin,  and K. W. West, Solid State Commun. {\bf 109}, 389 (1999);
K. B. Cooper, M. P. Lilly, J. P. Eisenstein, L. N. Pfeiffer, and K. W. West, Phys. Rev. B {\bf 60}, R11285 (1999).

\bibitem{kfs_prl} A. A. Koulakov, M. M. Fogler, and B. I. Shklovskii, Phys. Rev. Lett. {\bf 76}, 499 (1996).

\bibitem{kfs_prb} M. M. Fogler, A. A. Koulakov, and B. I. Shklovskii, Phys. Rev. B {\bf 54}, 1853 (1996).

\bibitem{moessner_chalker} R. Moessner and J. T. Chalker, Phys. Rev. B {\bf 54}, 5006 (1996).



\bibitem{edduncanky_stripe} E. H. Rezayi, K. Yang, and F. D. M. Haldane, Phys. Rev. Lett. {\bf 83}, 1219 (1999).

\bibitem{sma} S. M. Girvin, A. H. MacDonald, and P. M. Platzman, Phys. Rev. Lett. {\bf 54}, 581 (1985); S. M. Girvin, A. H. MacDonald, and P. M. Platzman,
Phys. Rev. B {\bf 33}, 2481 (1986). 



\bibitem{duncan_unpublished} F. D. M. Haldane, Phys. Rev. Lett. {\bf 107}, 116801 (2011). 

\bibitem{kunyang_aniso} R.-Z. Qiu, F. D. M. Haldane, Xin Wan, Kun Yang, and Su Yi, Phys. Rev. B {\bf 85}, 115308 (2012).

\bibitem{boyang} Bo Yang, Z. Papi\'c, E. H. Rezayi, R. N. Bhatt, and F. D. M. Haldane, Phys. Rev. B {\bf 85}, 165318 (2012).

\bibitem{xinwan} Hao Wang, Rajesh Narayanan, Xin Wan, and Fuchun Zhang, Phys. Rev. B {\bf 86}, 035122 (2012).

\bibitem{nematic}  L. Balents, Europhysics Letters {\bf 33}, 291 (1996); K. Musaelian and R. Joynt, Journal of Physics: Condensed Matter {\bf 8}, L105 (1996); E. Fradkin and S. A. Kivelson, Phys. Rev. B {\bf 59}, 8065 (1999);
M. M. Fogler, Europhysics Letters {\bf 66}, 572 (2004).

\bibitem{rh_pf}
E. H. Rezayi and F. D. M. Haldane, Phys. Rev. Lett. {\bf 84}, 4685 (2000).



\bibitem{readgreen} N. Read and D. Green, Phys. Rev. B {\bf 61}, 10267 (2000).

\bibitem{kamburov} Dobromir Kamburov, Yang Liu, Mansour Shayegan, Loren N. Pfeiffer, Kenneth W. West, Kirk W. Baldwin,  arXiv:1302.2540 (2013).

\bibitem{fubini} R. Roy, arXiv:1208.2055 (2012); E. Dobardzic, M.V. Milovanovic, and N. Regnault, arXiv:1303.7131 (2013).

\bibitem{mulligan} M. Mulligan, C. Nayak, and S. Kachru, Phys. Rev. B {\bf 82}, 085102 (2010). 

\bibitem{maciejko} J. Maciejko, B. Hsu, S. A. Kivelson, YeJe Park, S. L. Sondhi,  arXiv:1303.3041 (2013).


\end{thebibliography}
\end{document}